\documentclass[showpacs,aps,graphicx,twocolumn]{revtex4}
\usepackage{graphicx}
\usepackage{subfigure}
\begin{document}
\title{ Distilling single-photon entanglement from photon loss and decoherence}

\author{ Yu-Bo Sheng,$^{1,2}$\footnote{Email address:
shengyb@njupt.edu.cn} Lan Zhou,$^{2,3}$  }
\address{$^1$Institute of Signal Processing  Transmission, Nanjing
University of Posts and Telecommunications, Nanjing, 210003,  China}
\address{$^2$Key Lab of Broadband Wireless Communication and Sensor Network
 Technology, Nanjing University of Posts and Telecommunications, Ministry of
 Education, Nanjing, 210003, China}
\address{$^3$College of Mathematics \& Physics, Nanjing University of Posts and Telecommunications, Nanjing,
210003, China}


\date{\today}

\begin{abstract}
Single-photon entanglement may be the simplest type of entanglement but it is of vice importance
 in quantum communication. Here we present a practical  protocol for
distilling  the single-photon entanglement from both  photon loss and decoherence. With the help of some local single photons, the probability of single photon loss can be decreased and the less-entangled state can also be recovered to maximally entangled state simultaneously.
 It only requires some linear optical elements which makes it feasible in current experiment
condition. This protocol might find applications in current quantum communications based on the quantum repeaters.
\end{abstract}
 \maketitle


\section{introduction}
Quantum entanglement is the key source in current quantum information processing (QIP) \cite{book,rmp}. Most quantum communication protocols
such as quantum key distribution \cite{Ekert91,QSDC1}, teleportation \cite{teleportation,cteleportation1,cteleportation2}, quantum secret sharing \cite{QSS1,QSS2,QSS3}, quantum secure direct communication \cite{QSDC2,QSDC3}, and quantum state sharing \cite{QSTS1,QSTS2,QSTS3} all need the entanglement.
Single-photon entanglement (SPE) of the form $\frac{1}{\sqrt{2}}(|0\rangle_{A}|1\rangle_{B}+|1\rangle_{A}|0\rangle_{B})$ may be the simplest
type of entanglement. It means that the single photon may be in the location A, or in the location B with the same probability.
The single-photon entangled state can be prepared with the single-photon source and the 50:50 beam splitter (BS).
The early works for SPE are focused on the nolocality of the single photon \cite{single1,single2,single3}. In the recent ten years,
the SPE has been
proved to be a valuable resource for cryptography \cite{singleQKD1,singleQKD2},
state engineering \cite{engineering}, and tomography of states and
operations \cite{tomography1,tomography2}. Especially,  Duan \emph{ et al.} proposed a quantum repeater protocol using the single-photon
entanglement and cold atomic ensembles in 2001 \cite{DLCZ}. It is so called the DLCZ protocol. The quantum repeater protocols with SPE
were well developed by the group of Gisin \cite {singlerepeater1,singlerepeater2,singlerepeater3}. Recently, the entanglement  purification and entanglement
concentration protocols for SEM protocol were also realized and discussed \cite{singlepurification1,singlepurification2,shengqic,shengjosa1}, respectively. The extraction of information from the single particle using partial measurement were well discussed by Paraoanu \cite{partialmeasurement1,partialmeasurement2,partialmeasurement3}.

However, like any other quantum entanglement based on a two-photon pair or two-mode electromagnetic field, SPE cannot
avoid the environment noise.
The degree of entanglement
 between two distant sites normally decreases exponentially
with the length of the connecting channel  \cite{DLCZ}. Therefore, the photon loss is one of the main
obstacle for long-distance quantum communication. One of the powerful  ways to
overcome the difficulty associated with the exponential fidelity
decay is the photon noiseless  linear amplification (NLA), which was  proposed by Ralph and Lund \cite{amplification1}.
In 2010, the proposal for implementing Device-Independent-Quantum key distribution based on
a heralded qubit amplifier was proposed \cite{amplification2}. Xiang \emph{et al.} also realized the heralded NLA and
the distillation of entanglement in the same year \cite{amplification3}.  Curty and Moroder  discussed the heralded-qubit amplifier for practical
device-independent quantum key distribution \cite{amplification4}. Pitkanen also presented an efficient way for heralding photonic qubit signals using linear
optics devices \cite{amplification5}. Recently, Osorio \emph{et al.} reported their experimental result for heralded noiseless amplification based on single-photon sources
and linear optics  \cite{amplification6}. The experiment of heralded noiseless amplification of a photon polarization qubit was also reported
 recently \cite{amplification8}. Inspired by the previous works, Zhang \emph{et al.} proposed an efficient way for protecting
single-photon entanglement from photon loss using NLA  \cite{amplification7}. In their protocol, based on the model of photon loss, the maximally entangled state
$\frac{1}{\sqrt{2}}(|0\rangle_{A}|1\rangle_{B}+|1\rangle_{A}|0\rangle_{B})$ may become the mixed state of the form
\begin{eqnarray}
\rho_{AB}=\eta|\Phi\rangle_{AB}\langle\Phi|+(1-\eta)|vac\rangle\langle vac|.\label{model}
\end{eqnarray}
Here $|\Phi\rangle_{AB}=\frac{1}{\sqrt{2}}(|0\rangle_{A}|1\rangle_{B}+|1\rangle_{A}|0\rangle_{B})$ and $|vac\rangle$ is the vacuum state, which
means that the photon is completely lost with the probability of $1-\eta$.

Actually, during the practical transmission,
though the sing-photon entanglement is not completely lost, the parties still cannot obtain the maximally entangled state. The reason is that the
single photon can be partially lost which can make $|\Phi\rangle_{AB}$ degrade to
$|\Psi\rangle_{AB}=\alpha|1\rangle_{A}|0\rangle_{B}+\beta|0\rangle_{A}|1\rangle_{B}$. Here  $|\alpha|^{2}+|\beta|^{2}=1$. Therefore, in order to obtain
the maximally entangled state, the parties not only need to increase the probability of the SPE, but also need
to recover the less-entangled state into the maximally entangled state.

Entanglement concentration provides us a good way for distilling the maximally entangled state from
the less-entangled state \cite{C.H.Bennett2,zhao1,Yamamoto1,swapping1,swapping2,shengpra2,shengpra3,dengconcentration,wangchuan1,wangchuan2,
zhounoon,shengwstate,review}. The concept of the entanglement concentration for two-particle less-entangled state was proposed by Bennett \emph{et al.} in 1996 \cite{C.H.Bennett2}. The entanglement concentration protocol (ECP)
based on the linear optics for the photon pair encoded in the polarization degree of freedom was proposed by Zhao \emph{et al.} and Yamamoto \emph{et al.}
independently \cite{zhao1,Yamamoto1}. Later, their ECPs ware developed with the help of cross-Kerr nonlinearity \cite{shengpra2,shengpra3,dengconcentration}.

\begin{figure}[!h]
\begin{center}
\includegraphics[width=7cm,angle=0]{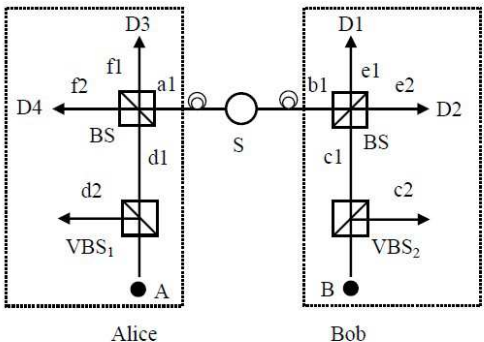}
\caption{A schematic drawing of the ELACP for single-photon entangled state. Two different VBSs are used here.}
\end{center}
\end{figure}
Therefore, inspired by the practical environment noise and the previous works about NLA and ECPs,
in this paper, we will describe an practical way for distilling the single-photon entangled state from both the photon loss and decoherence. It contains
both linear amplification and concentration. Therefore,
this entanglement linear amplification and concentration protocol (ELACP) can both amplify the fidelity of the entangled state and distill the maximally entangled state from the less-entangled state.
Interestingly, with the help of two local single photons,  by adjusting the different transmission
coefficient of two variable fiber beam splitters (VBSs), the whole task can be achieved with some success probability.
Moreover, the similar setup described in Refs. \cite{amplification6,amplification7} can be used to complete this protocol and only two different VBSs are required, which make this protocol
be useful for current quantum communication.

This paper is organized as follows: In Sec. II, we first briefly described the basic model of this ELACP. In Sec. III, we provide our
numerical simulation and discussion. In Sec. IV, we discuss the practical realization in current experiment.  In Sec. V, we make a conclusion.

\section{linear amplification and concentration}
In this section, we start to explain our EALCP model. As shown in Fig. 1,
the single-photon source (S) emits one pair of maximally entangled state and
distributes it to Alice and Bob. However, the lossy channel will degrade it and make
it become
\begin{eqnarray}
\rho_{AB}=\eta|\Psi\rangle_{a_{1}b_{1}}\langle\Psi|+(1-\eta)|vac\rangle\langle vac|.\label{initial}
\end{eqnarray}
Here $|\Psi\rangle_{a1b1}$ is a less-entangled state with $|\Psi\rangle_{a1b1}=\alpha|1\rangle_{a1}|0\rangle_{b1}+\beta|0\rangle_{a1}|1\rangle_{b1}$,
where $|\alpha|^{2}+|\beta|^{2}=1$. From Eq. (\ref{initial}), after the noisy channel,
the single photon may be completely lost and become a vacuum sate $|vac\rangle$ with the probability of $1-\eta$. It may
be partially lost and become a less-entangled state $|\Psi\rangle_{a1b1}$. Therefore, the task of this protocol is not only to
amplify the single-photon entangled state, but also to convert  the less-entangled state to the maximally entangled state.

From Fig. 1, two variable fiber beam splitters (VBSs) are used here. Different from Ref. \cite{amplification7}, the transmission of two
VBSs are different. The transmission of  VBS$_{1}$  and VBS$_{2}$ is t$_{1}$ and  t$_{2}$, respectively.
They will produce the entangled state between different spatial modes as
\begin{eqnarray}
|1\rangle_{A}\rightarrow\sqrt{t_{1}}|1\rangle_{d1}|0\rangle_{d2}+\sqrt{1-t_{1}}|0\rangle_{d1}|1\rangle_{d2},\\
|1\rangle_{B}\rightarrow\sqrt{t_{2}}|1\rangle_{c1}|0\rangle_{c2}+\sqrt{1-t_{2}}|0\rangle_{c1}|1\rangle_{c2}.
\end{eqnarray}

\begin{figure}[!h]
\begin{center}
\includegraphics[width=7cm,angle=0]{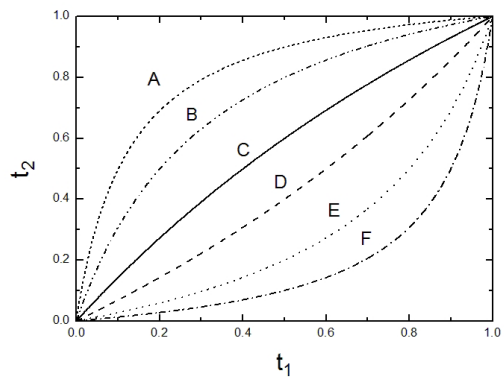}
\caption{The transmission coefficient t$_{2}$ is altered with the t$_{1}$ to satisfy the concentration condition as
shown in Eq. (\ref{condition1}). Curve A: $\alpha^{2}$=0.1.
Curve B: $\alpha^{2}$=0.2. Curve C: $\alpha^{2}$=0.4. Curve D:$\alpha^{2}$=0.5. Curve E: $\alpha^{2}$=0.8. Curve F: $\alpha^{2}$=0.9.}
\end{center}
\end{figure}
Here the $|1\rangle_{A}$ and $|1\rangle_{B}$ are two single photons prepared by Alice and Bob, respectively.
Therefore, the two single photons combined with the $\rho_{AB}$ can be described as:
with the probability of $\eta$, it is in the state $|\Psi\rangle_{a_{1}b_{1}}\otimes|1\rangle_{A}\otimes|1\rangle_{B}$, with the
probability of $1-\eta$, it is in the state $|vac\rangle\otimes|1\rangle_{A}\otimes|1\rangle_{B}$.
We first describe the $|\Psi\rangle_{a_{1}b_{1}}\otimes|1\rangle_{A}\otimes|1\rangle_{B}$. It  evolves as
\begin{eqnarray}
&&|\Psi\rangle_{a_{1}b_{1}}\otimes|1\rangle_{A}\otimes|1\rangle_{B}=(\alpha|1\rangle_{a1}|0\rangle_{b1}+\beta|0\rangle_{a1}|1\rangle_{b1})\nonumber\\
&\otimes&|1\rangle_{A}\otimes|1\rangle_{B}\rightarrow(\alpha|1\rangle_{a1}|0\rangle_{b1}+\beta|0\rangle_{a1}|1\rangle_{b1})\nonumber\\
&\otimes&(\sqrt{t_{1}}|1\rangle_{d1}|0\rangle_{d2}+\sqrt{1-t_{1}}|0\rangle_{d1}|1\rangle_{d2})\nonumber\\
&\otimes&(\sqrt{t_{2}}|1\rangle_{c1}|0\rangle_{c2}+\sqrt{1-t_{2}}|0\rangle_{c1}|1\rangle_{c2})\nonumber\\
&=&\alpha\sqrt{t_{1}t_{2}}|1\rangle_{a1}|0\rangle_{b1}|1\rangle_{d1}|0\rangle_{d2}|1\rangle_{c1}|0\rangle_{c2}\nonumber\\
&+&\alpha\sqrt{t_{1}(1-t_{2})}|1\rangle_{a1}|0\rangle_{b1}|1\rangle_{d1}|0\rangle_{d2}|0\rangle_{c1}|1\rangle_{c2}\nonumber\\
&+&\alpha\sqrt{t_{2}(1-t_{1})}|1\rangle_{a1}|0\rangle_{b1}|0\rangle_{d1}|1\rangle_{d2}|1\rangle_{c1}|0\rangle_{c2}\nonumber\\
&+&\alpha\sqrt{(1-t_{1})(1-t_{2})}|1\rangle_{a1}|0\rangle_{b1}|0\rangle_{d1}|1\rangle_{d2}|0\rangle_{c1}|1\rangle_{c2}\nonumber\\
&+&\beta\sqrt{t_{1}t_{2}}|0\rangle_{a1}|1\rangle_{b1}|1\rangle_{d1}|0\rangle_{d2}|1\rangle_{c1}|0\rangle_{c2}\nonumber\\
&+&\beta\sqrt{t_{1}(1-t_{2})}|0\rangle_{a1}|1\rangle_{b1}|1\rangle_{d1}|0\rangle_{d2}|0\rangle_{c1}|1\rangle_{c2}\nonumber\\
&+&\beta\sqrt{t_{2}(1-t_{1})}|0\rangle_{a1}|1\rangle_{b1}|0\rangle_{d1}|1\rangle_{d2}|1\rangle_{c1}|0\rangle_{c2}\nonumber\\
&+&\beta\sqrt{(1-t_{1})(1-t_{2})}|0\rangle_{a1}|1\rangle_{b1}|0\rangle_{d1}|1\rangle_{d2}|0\rangle_{c1}|1\rangle_{c2}.\label{tuidao1}
\end{eqnarray}
In order to realize the protocol, both Alice and Bob should detect one and only one photon. From Eq. (\ref{tuidao1}), terms
$|1\rangle_{a1}|0\rangle_{b1}|1\rangle_{d1}|0\rangle_{d2}|1\rangle_{c1}|0\rangle_{c2}$, $|1\rangle_{a1}|0\rangle_{b1}|1\rangle_{d1}|0\rangle_{d2}|0\rangle_{c1}|1\rangle_{c2}$,$|1\rangle_{a1}|0\rangle_{b1}|0\rangle_{d1}|1\rangle_{d2}|0\rangle_{c1}|1\rangle_{c2}$,
$|0\rangle_{a1}|1\rangle_{b1}|1\rangle_{d1}|0\rangle_{d2}|1\rangle_{c1}|0\rangle_{c2}$, $|0\rangle_{a1}|1\rangle_{b1}|0\rangle_{d1}|1\rangle_{d2}|1\rangle_{c1}|0\rangle_{c2}$ and $|0\rangle_{a1}|1\rangle_{b1}|0\rangle_{d1}|1\rangle_{d2}|0\rangle_{c1}|1\rangle_{c2}$
cannot lead the Alice and Bob both only detect one photon. Therefore, by choosing the
cases that both Alice and Bob only detect one photon, Eq. (\ref{tuidao1}) will become
\begin{eqnarray}
|\Psi_{1}\rangle&=&\alpha\sqrt{t_{2}(1-t_{1})}|1\rangle_{a1}|0\rangle_{b1}|0\rangle_{d1}|1\rangle_{d2}|1\rangle_{c1}|0\rangle_{c2}\nonumber\\
&+&\beta\sqrt{t_{1}(1-t_{2})}|0\rangle_{a1}|1\rangle_{b1}|1\rangle_{d1}|0\rangle_{d2}|0\rangle_{c1}|1\rangle_{c2}.\label{collapse1}
\end{eqnarray}
Certainly, $|vac\rangle\otimes|1\rangle_{A}\otimes|1\rangle_{B}$ can also lead Alice and Bob both obtain one photon.
In detail,
\begin{eqnarray}
&&|vac\rangle\otimes|1\rangle_{A}\otimes|1\rangle_{B}\nonumber\\
&=&|vac\rangle\otimes(\sqrt{t_{1}}|1\rangle_{d1}|0\rangle_{d2}+\sqrt{1-t_{1}}|0\rangle_{d1}|1\rangle_{d2})\nonumber\\
&\otimes&(\sqrt{t_{2}}|1\rangle_{c1}|0\rangle_{c2}+\sqrt{1-t_{2}}|0\rangle_{c1}|1\rangle_{c2})\nonumber\\
&=&|vac\rangle\otimes(\sqrt{t_{1}t_{2}}|1\rangle_{d1}|0\rangle_{d2}|1\rangle_{c1}|0\rangle_{c2}\nonumber\\
&+&\sqrt{(1-t_{1})t_{2}}|0\rangle_{d1}|1\rangle_{d2}|1\rangle_{c1}|0\rangle_{c2}\nonumber\\
&+&\sqrt{t_{1}(1-t_{2})}|1\rangle_{d1}|0\rangle_{d2}|0\rangle_{c1}|1\rangle_{c2}\nonumber\\
&+&\sqrt{(1-t_{1})(1-t_{2})}|0\rangle_{d1}|1\rangle_{d2}|0\rangle_{c1}|1\rangle_{c2}).\label{tuidao2}
\end{eqnarray}
From Eq. (\ref{tuidao2}), they will obtain
\begin{eqnarray}
|\Psi_{2}\rangle=(\sqrt{t_{1}t_{2}}|1\rangle_{d1}|0\rangle_{d2}|1\rangle_{c1}|0\rangle_{c2},\label{collapse2}
\end{eqnarray}
if Alice and Bob both contain one photon. The total success probability is
\begin{eqnarray}
P=\eta(\alpha^{2}t_{2}+\beta^{2}t_{1})+t_{1}t_{2}-2\eta t_{1}t_{2}. \label{success}
\end{eqnarray}
In Fig. 1, the beam splitters (BSs) are used to convert
\begin{eqnarray}
|1\rangle_{a1}\rightarrow\frac{1}{\sqrt{2}}(|1\rangle_{f1}+|1\rangle_{f2}),\nonumber\\
|1\rangle_{d1}\rightarrow\frac{1}{\sqrt{2}}(|1\rangle_{f1}-|1\rangle_{f2}).\nonumber\\
|1\rangle_{b1}\rightarrow\frac{1}{\sqrt{2}}(|1\rangle_{e1}+|1\rangle_{e2}),\nonumber\\
|1\rangle_{c1}\rightarrow\frac{1}{\sqrt{2}}(|1\rangle_{e1}-|1\rangle_{e2}).
\end{eqnarray}
\begin{figure}[!h]
\begin{center}
\includegraphics[width=7cm,angle=0]{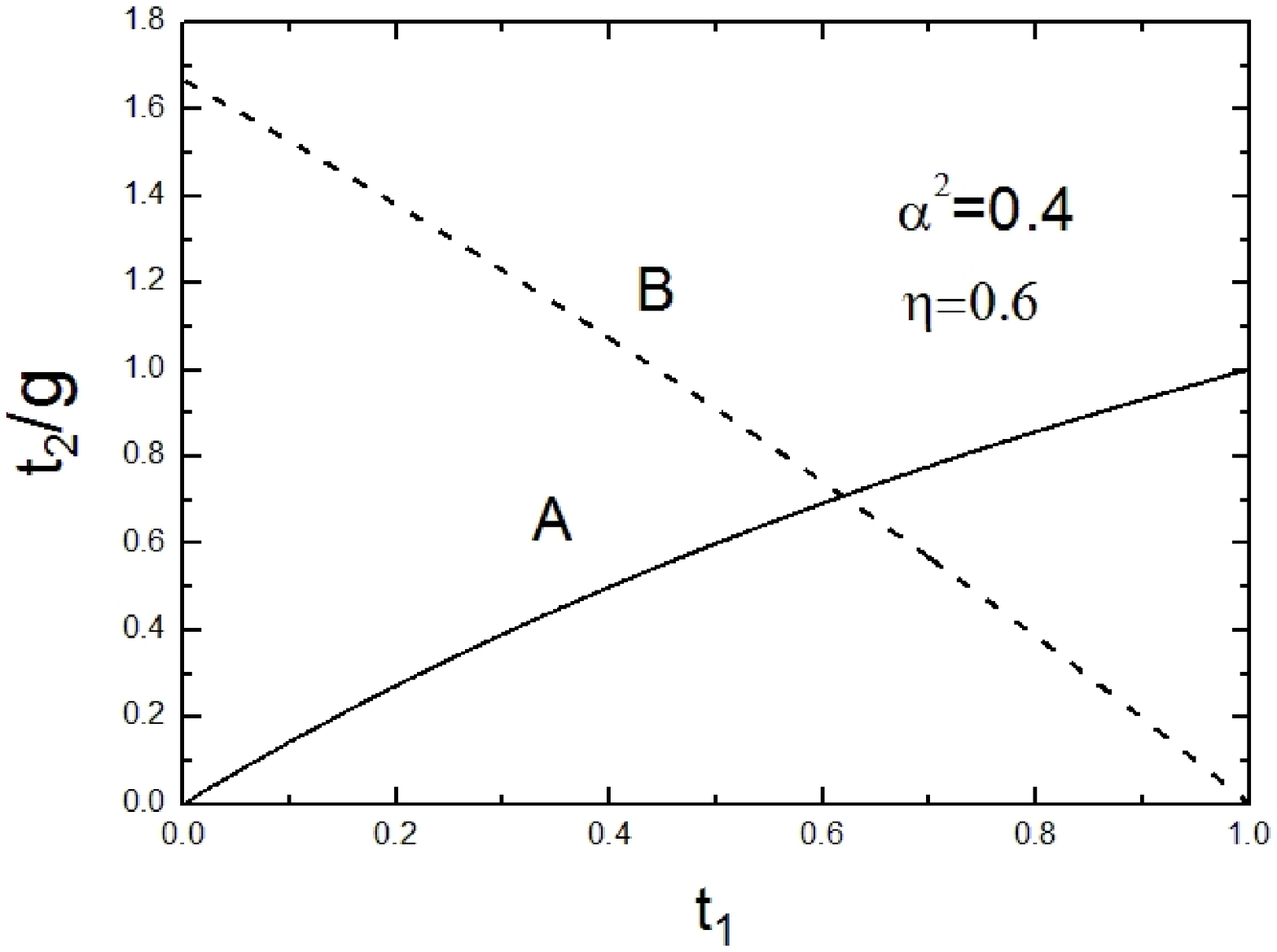}
\caption{The transmission coefficient t$_{2}$ (Curve A) and the amplification factor g (Curve B) are altered with t$_{1}$, with $\alpha^{2}$=0.4. Here we choose $\eta=0.6$.}
\end{center}
\end{figure}

\begin{figure}[!h]
\begin{center}
\includegraphics[width=7cm,angle=0]{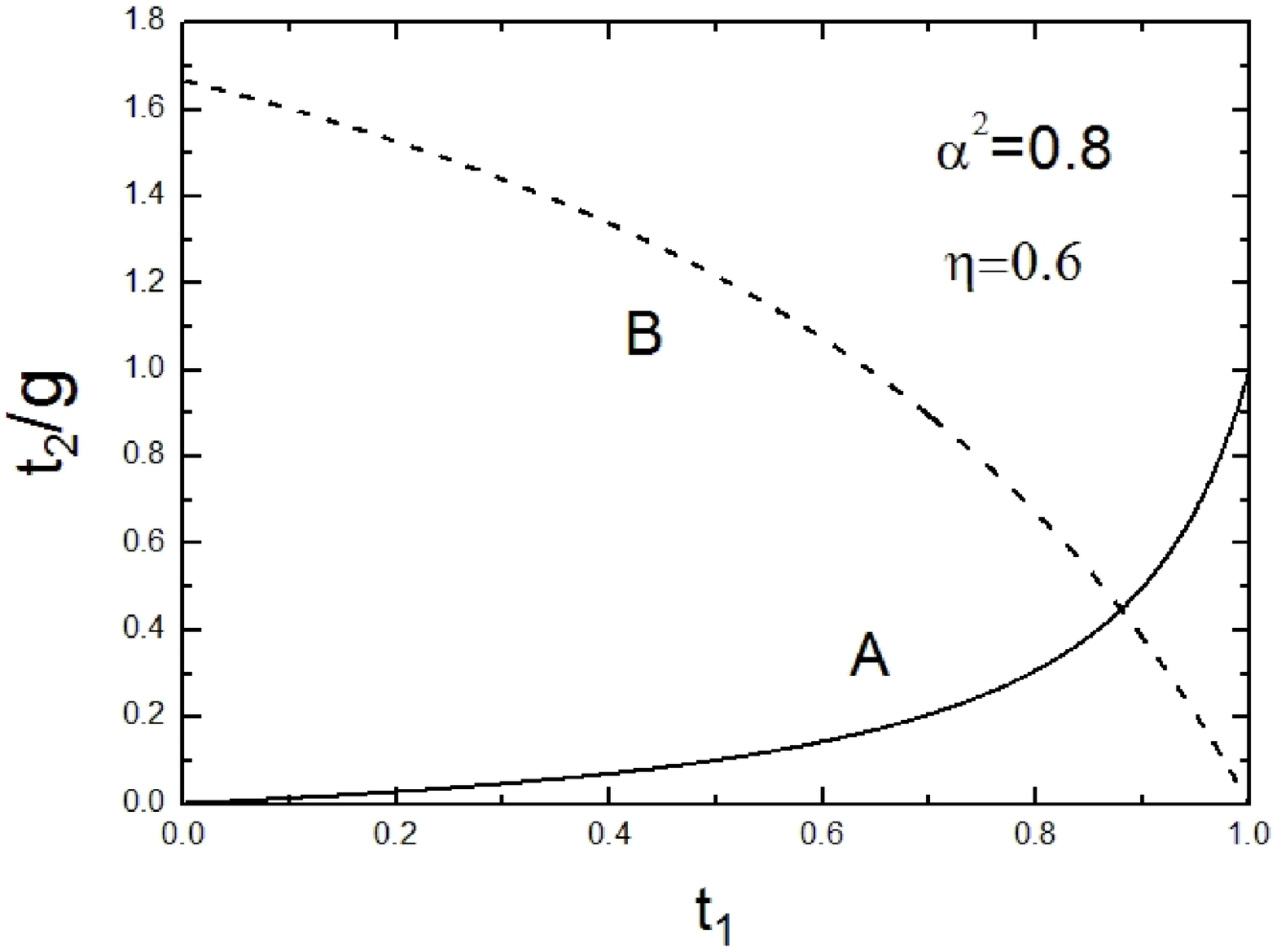}
\caption{The transmission coefficient t$_{2}$ (Curve A) and the amplification factor g (Curve B) are altered with t$_{1}$, with $\alpha^{2}$=0.8. Here we choose $\eta=0.6$.}
\end{center}
\end{figure}
Therefore, if the single-photon detectors D$_{1}$D$_{3}$ or D$_{2}$D$_{4}$ register, $|\Psi_{1}\rangle$ will become
\begin{eqnarray}
|\Psi_{2}\rangle'=&&\alpha\sqrt{t_{2}(1-t_{1})}|1\rangle_{a1}|0\rangle_{b1}\nonumber\\
&+&\beta\sqrt{t_{1}(1-t_{2})}|0\rangle_{a1}|1\rangle_{b1}.
\end{eqnarray}
On the other hand, if  D$_{2}$D$_{3}$ or D$_{1}$D$_{4}$ register, $|\Psi_{1}\rangle$ will become
\begin{eqnarray}
|\Psi_{2}\rangle''&=&\alpha\sqrt{t_{2}(1-t_{1})}|1\rangle_{a1}|0\rangle_{b1}\nonumber\\
&-&\beta\sqrt{t_{1}(1-t_{2})}|0\rangle_{a1}|1\rangle_{b1}.
\end{eqnarray}
One of the parties only needs to perform a phase-flip operation to convert it to $|\Psi_{2}\rangle'$.
 $|\Psi_{2}\rangle$ always becomes the vacuum state after the two photons being detected.

Therefore, after the two photons being detected, the initial state becomes
\begin{eqnarray}
\rho'_{AB}=\eta'|\Psi'_{2}\rangle\langle\Psi'_{2}|+(1-\eta')|vac\rangle\langle vac|.
\end{eqnarray}
Here
\begin{eqnarray}
\eta'=\frac{\eta(\alpha^{2}t_{2}+\beta^{2}t_{1}-t_{1}t_{2})}{\eta(\alpha^{2}t_{2}+\beta^{2}t_{1})+t_{1}t_{2}-2\eta t_{1}t_{2}}.
\end{eqnarray}

In order to obtain the maximally entangled state, they should let
\begin{eqnarray}
\frac{\alpha^{2}}{\beta^{2}}=\frac{t_{1}(1-t_{2})}{{t_{2}(1-t_{1})}}.\label{condition1}
\end{eqnarray}

Moreover, in order to realize the linear amplification, they should make $\eta'>\eta$. If we denote the ampliation factor $g\equiv\frac{\eta'}{\eta}$, we can obtain
\begin{eqnarray}
g=\frac{\alpha^{2}t_{2}+\beta^{2}t_{1}-t_{1}t_{2}}{\eta(\alpha^{2}t_{2}+\beta^{2}t_{1})+t_{1}t_{2}-2\eta t_{1}t_{2}}.\label{condition2}
\end{eqnarray}

\section{Numerical simulation and discussion}
Thus far, we have briefly explained the EALCP. From above discussion, we should require
two different VBSs, while in Ref.\cite{amplification7}, two similar VBSs are used. In Eq. (\ref{condition1}), we should adjust
the coefficients t$_{1}$ and t$_{2}$ to obtain the maximally entangled state, according to the initial coefficients $\alpha$ and $\beta$. Certainly, if we choose a special case that $\alpha^{2}=\beta^{2}=\frac{1}{2}$,
 we can obviously obtain t$_{1}$=t$_{2}$ $\equiv$t. Moreover, according to the  Eq. (\ref{condition2}), we can get $g=\frac{1-t}{\eta+t-2\eta t}$. In this way, our results is consist with the Ref.\cite{amplification7}.  The relationship between t$_1$ and t$_2$ is shown in Fig. 2.
We calculated the t$_{2}$ alters  with t$_{1}$ under different $\alpha^{2}$. In Fig. 2, Curve A is corresponding to  $\alpha^{2}=0.1$.
 Curve B is corresponding to $\alpha^{2}=0.2$.  Curve C is corresponding to  $\alpha^{2}=0.4$.
  Curve D is corresponding to $\alpha^{2}=0.6$.   Curve E is corresponding to  $\alpha^{2}=0.8$, and
  Curve F is corresponding to  $\alpha^{2}=0.9$.  From Eq. (\ref{condition1}), if $\alpha=\beta$, we can obtain that
  t$_{1}$=t$_{2}$, which is same as Ref. \cite{amplification7}. The relationship  described in Fig. 2 only realizes the function of entanglement concentration. In order to
  increase the fidelity of the entangled state, we should let $\eta'>\eta$.
\begin{figure}[!h]
\begin{center}
\includegraphics[width=7cm,angle=0]{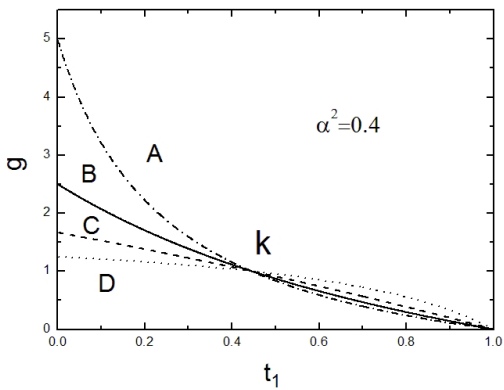}
\caption{The amplification factor $g$ is altered with the transmission coefficient t$_{1}$ with $\alpha^{2}=0.4$.
Curve A: $\eta=0.2$. Curve B: $\eta=0.4$. Curve C:  $\eta=0.6$. Curve D: $\eta=0.8$. All four curves pass through the same point $K$ with $g=1$.}
\end{center}
\end{figure}

\begin{figure}[!h]
\begin{center}
\includegraphics[width=7cm,angle=0]{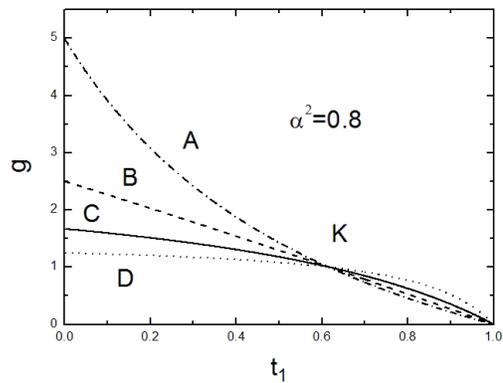}
\caption{The amplification factor $g$ is altered with the transmission coefficient t$_{1}$ with $\alpha^{2}=0.8$.
Curve A: $\eta=0.2$. Curve B: $\eta=0.4$. Curve C:  $\eta=0.6$. Curve D: $\eta=0.8$. All four curves pass through the same point $K$ with $g=1$.
The numerical simulation of the limitation $g$ shows that it is equal to $\frac{1}{\eta}$, which is consist to the theoretical derivation. }
\end{center}
\end{figure}
In Fig. 3 and Fig. 4, we  both calculated the t$_{2}$ and amplification factor $g$ alters with the t$_{1}$ when $\eta_{0}=0.4$, $\alpha^{2}=0.4$ and $\alpha^{2}=0.8$, respectively. In Fig. 3, Curve A is the t$_{2}$ alters with the coefficient t$_{1}$ and the Curve B is the g alters with  t$_{1}$. Therefore,
in order to realize the amplification, we should require $t_{1} \in (0,0.44)$, and  $t_{2} \in (0,0.54)$ when  $\eta=0.6$, $\alpha^{2}=0.4$.
On the other hand, In Fig. 4, when  $\eta=0.6$, $\alpha^{2}=0.8$, we get $t_{1} \in (0,0.63)$ and  $t_{1} \in (0,0.15)$.
We also calculated the $g$ alters with the different initial probability $\eta$ and $\alpha^{2}$. In Fig. 5 and Fig. 6, Curve A is
relationship between t$_{1}$ and $g$ when $\eta=0.2$.  Curve B is $\eta=0.4$. Curve C is $\eta=0.6$ and Curve D is $\eta=0.8$. Interestingly,
as shown in Fig. 5 and Fig. 6, all four curves pass through the same point $K$, which corresponding to $\eta=1$. Actually, according to Eq. (\ref{condition2}),
if  $\eta=1$, we can get $\frac{\alpha^{2}t_{2}+\beta^{2}t_{1}}{2t_{1}t_{2}}=1$. It is shown that the initial coefficient $\eta$ disappears. This is the reason
that all curves pass through the same point $K$.  From both Figs. 5  and 6, another interesting case is that the limitation of $g$ seems to
be a constant. It changes with initial $\eta$ and has nothing to do with the initial $\alpha$ and $\beta$. From Eq. (\ref{condition1}) and Fig. 2, $t_{2}\rightarrow0$ when $t_{1}\rightarrow0$. In Eq. (\ref{condition2}), term $t_{1}t_{2}$ is the second-order infinitesimal when $t_{2}\rightarrow0$ and $t_{1}\rightarrow0$, which can be omitted. Therefore, we can calculate the limitation of $g$ as
\begin{eqnarray}
\lim_{t_{1},t_{2}\rightarrow0}g=\frac{\alpha^{2}t_{2}+\beta^{2}t_{1}}{\eta(\alpha^{2}t_{2}+\beta^{2}t_{1})}=\frac{1}{\eta}.\label{lim}
\end{eqnarray}
The numerical simulation shown in Fig. 5 and Fig.  6 is consist with the Eq. (\ref{lim}). It is easy to explain the limitation $g$ in  Eq. (\ref{lim}),
for the upper limit of both $\eta$ and $\eta'$  is 1, according to Eq. (\ref{initial}). If the initial probability $\eta$ is 0.2,
the maximally value of $g$ must be 5. In this way, the limitation of $g$ is nothing to do with the coefficient $\alpha$. We should point out that
if both $t_{2}\rightarrow0$ and $t_{1}\rightarrow0$, from Eq. (\ref{success}), the total success probability is also $P\rightarrow 0$. It reveals that it is a
trade-off between the success probability $P$ and amplification $g$.

\section{Possible experimental realization}
\begin{figure}[!h]
\begin{center}
\includegraphics[width=9cm,angle=0]{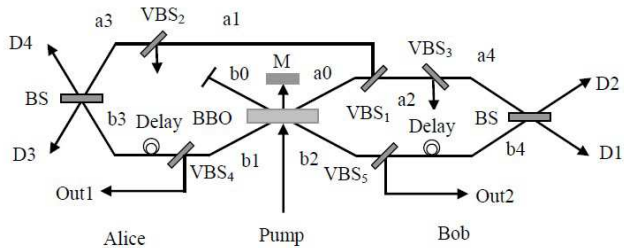}
\caption{A schematic drawing of the possible experimental realization of ELACP for single-photon entangled state.}
\end{center}
\end{figure}

In this section, we will briefly discuss the possible realization in experiment. The main challenge for this ELACP is to ensure the input and auxiliary photons are indistinguishable in every degree of freedom. They are essentially
the Hong-Ou-Mandel (HOM) interference on two BSs. As shown in Fig. 7, the pump
pulse of ultraviolet light passes through a $\beta$-barium borate
crystal (BBO). Then a correlated pair of photons will be generated
with the probability $p$ in the spatial modes $a_{0}$ and $b_{0}$. The pump
pulse of ultraviolet light will be reflected by the mirror (M) and pass through the
BBO crystal in a second time, which can generate another correlated pair in the spatial mode  $b_{1}$ and $b_{2}$. The
position of the mirror can be adjusted. The Delay line can also be used to ensure the photons arriving at the
two BSs simultaneously. The photon in the spatial mode   $b_{0}$ is blocked. The photon in the
spatial mode $a_{0}$ is used to prepare the initial state as shown in Eq. (\ref{initial}). The VBS$_{1}$ is to generate
the less-entangled state $|\Psi\rangle$, similar to Ref. \cite{amplification7}. The coefficient $\alpha$ is decided by the transmission of VBS$_{1}$.
The VBS$_{2}$ and VBS$_{3}$ with the same transmission are used to prepare the local bosonic lossy channels.
The transmission coefficients of  VBS$_{2}$ and VBS$_{3}$ is $T_{0}=\eta$. The two auxiliary photons  are prepared in the spatial modes
b$_{1}$ and b$_{2}$, respectively. According to the initial $\eta$ and $\alpha$,
we should  adjust the transmission of VBS$_{4}$ and VBS$_{5}$ to satisfy the relationship in Eqs. (\ref{condition1}) and
(\ref{condition2}), for its realization in the experiment. On the other hand, the phase fluctuations and multi-photon
cases may disturb the experiment. Fortunately, as pointed out by Refs. \cite{purification1,purification2}, the good overlap on the PBS and phase stability have been achieved in
current experimental condition. The second problem can be solved by four-mode cases, similar as Ref.\cite{purification2}. That is the spatial modes
b$_{0}$,  D$_{1}$, D$_{3}$ and Out1, or b$_{0}$,  D$_{1}$, D$_{3}$ and Out2 all exactly contain one photon.

\section{conclusion}
In conclusion, we have a practical way to distill the SPE from photon loss and decoherence.
By adjusting the transmission coefficients t$_{1}$ and t$_{2}$, one can
both realize the amplification and concentration of the single-photon entangled state.
 We discussed the numerical condition of amplification
factor $g$ changed with the initial parameters $\alpha$ and $\eta$.
Interestingly, the limitation $g$  only depends on the  $\eta$ and it has nothing to do with $\alpha$.
We also discussed its possible realization in current experimental condition. Compared with the protocol of SPE
amplification, this protocol is more practical. As the SEM is of vice importance in current long-distance
quantum communication, this protocol is useful.

\section*{ACKNOWLEDGEMENTS}
This work was supported by the National Natural Science Foundation
of China under Grant No. 11104159,  Open Research
Fund Program of the State Key Laboratory of
Low-Dimensional Quantum Physics, Tsinghua University,
Open Research Fund Program of National Laboratory of Solid State Microstructures under Grant No. M25022, Nanjing University, the open research fund of Key Lab of Broadband Wireless Communication and Sensor Network Technology, Nanjing University of Posts and Telecommunications, Ministry of Education (No. NYKL201303)
 and A Project
Funded by the Priority Academic Program Development of Jiangsu
Higher Education Institutions.

\end{document}